%% file: paper.tex
\documentclass[conference,10pt]{IEEEtran}
\IEEEoverridecommandlockouts

\usepackage{amsfonts}
\usepackage{amsbsy}
\usepackage{amssymb}
\usepackage{amscd}
\usepackage[cmex10]{amsmath}
\usepackage{float}
\usepackage{graphicx}
\usepackage{algorithm}
\usepackage[noend]{algpseudocode}

\input{code}

\begin{document} 
\title{Linear Transceiver Optimization in Multicell MIMO Based on the Generalized Benders Decomposition
\thanks{\copyright~2015 IEEE. Personal use of this material is permitted. Permission from IEEE must be obtained for all other uses, in any current or future media, including reprinting/republishing this material for advertising or promotional purposes, creating new collective works, for resale or redistribution to servers or lists, or reuse of any copyrighted component of this work in other works.}
}%
\author{\IEEEauthorblockN{Rami Mochaourab and Mats Bengtsson}
\IEEEauthorblockA{Signal Processing Department, ACCESS Linnaeus Centre, KTH Royal Institute of Technology\\ E-mail: \{rami.mochaourab, mats.bengtsson\}@ee.kth.se.}}
\maketitle
\input{sections/abstract}

%
\IEEEpeerreviewmaketitle
\input{sections/introduction}
\input{sections/system_model}
\input{sections/problem_reformulation}
\input{sections/upper_bound}
\input{sections/simulations}
\input{sections/conclusions}
\input{sections/ack}
\input{sections/appendix}

\bibliographystyle{IEEEtran}
\bibliography{references}

\end{document}

%% file: code.tex
\newtheorem{theorem}{Theorem}
\newtheorem{lemma}{Lemma}
\newtheorem{proposition}{Proposition}

\newcommand{\mat}[1]{\boldsymbol{\mathrm{#1}}}
\newcommand{\tr}[1]{\text{tr}\left( #1\right)}

\newcommand{\pp}[1]{{\left( #1 \right)}}

\newcommand{\br}[1]{{\left\{ #1 \right\}}}
\newcommand{\norm}[1]{{ \left\Vert #1 \right\Vert }}
\newcommand{\abs}[1]{{ \left| #1 \right| }}

\newcommand{\snorm}[1]{{ \left\Vert #1 \right\Vert^2_{\text{{F}}} }}

\DeclareMathOperator*{\argmax}{arg\,max}
\DeclareMathOperator*{\maximize}{\text{max}}
\DeclareMathOperator*{\minimize}{\text{min}}

\def\lin{{\mathrm{lin} }}
\def\lblin{{\mathrm{lin-lb} }}
\def\evat{{\Big|}}
\def\H{{{\dag}}} 
\def\bH{{\mat{{H}}}} 
\def\bI{{\mat{{I}}}} 
\def\bZ{{\mat{{Z}}}} 
\def\bX{{\mat{{X}}}} 
\def\bY{{\mat{{Y}}}} 
\def\bz{{\mat{{z}}}} 
\newcommand{\collec}[1]{\br{{#1}}}

\def\setX{{\mathcal{X}}} 
\def\setU{{\mathcal{U}}} 
\def\setV{{\mathcal{V}}} 
\def\setK{{\mathcal{K}}} 

%% file: sections/abstract.tex
\begin{abstract}
We study the maximum sum rate optimization problem in the multiple-input multiple-output interfering broadcast channel. The multiple-antenna transmitters and receivers are assumed to have perfect channel state information. In this setting, finding the optimal linear transceiver design is an NP-hard problem. We show that a reformulation of the problem renders the application of generalized Benders decomposition suitable. The decomposition provides us with an optimization structure which we exploit to apply two different optimization approaches. While one approach is guaranteed to converge to a local optimum of the original problem, the other approach hinges on techniques which can be promising for devising a global optimization method.    
\end{abstract}

%% file: sections/introduction.tex
\section{Introduction}
Finding the maximum sum rate operating point in the multiple-input multiple-output (MIMO) interference channel is an open problem. It is proven that for the case of perfect channel state information at the transmitters and the receivers, the problem is NP-hard even for the single receive antennas case \cite{Liu2011}. Seeking the optimal solution is of paramount importance when evaluating low complexity distributed schemes such as in \cite{Scutari2009a,Schmidt2013}.

An approach to simplify the problem is to characterize the set of necessary transmission strategies for Pareto optimal operation for each user independently \cite{Park2012}. However, the parametrization in \cite{Park2012} reveals that the search space remains huge for finding the jointly sum rate optimal transmission strategy. Another approach in \cite{Bengtsson2012} parameterizes the necessarily optimal transmission strategies by Lagrangian multipliers and shows an uplink-downlink duality in the setting. Such characterization is useful for devising heuristic algorithms which can efficiently update the Lagrangian multiplies. 

Alternating optimization methods which reach local optima of the sum rate optimization problem are reported in \cite{Negro2010a,Shi2011} and recently in \cite{Brandt2014} for the case of hardware impairments at the transmitters and receivers. The mentioned approaches rely on optimization of the weighted minimum mean square error (MMSE) by successive optimization of the transmit covariance matrices, receive covariance matrices as well as a weighting matrices. Such methods can be implemented in a distributed fashion whenever it is possible to exchange the weighting matrices between the transmitters and receivers.

A technique which helps in finding global solutions in problems with special couplings between the optimization variables is the generalized Benders decomposition. This decomposition method has been applied for control problems with bilinear inequality constraints in \cite{Floudas90aglobal} and generalized to bilinear matrix inequality constraints in \cite{Beran1997}. 

We show that the maximum sum rate problem can be cast in a form with bilinear matrix constraints suitable for the generalized Benders decomposition. In doing so, we reveal a structured way to approach the solution of the NP-hard optimization problem. Consequently, we provide two optimization approaches. The first approach is based on the work in \cite{Floudas1989, Geoffrion72} which leads to a local optimum of the original problem. The second approach utilizes a novel approximation on the upper bound of the sum rate. Although convergence to a stationary point of the original problem is not verified in the second approach, simulation results illustrate its efficiency.

\emph{Outline:} After describing the system model and formulating the sum rate maximization problem in Section~\ref{sec:sys_model}, we make the necessary problem reformulation to cast the problem using generalized Benders decomposition in Section~\ref{sec:Bender}. Exploiting the decomposition, two approaches to deal with the original problem are provided in Section~\ref{sec:approaches}. Discussion of the results using numerical examples are given in Section~\ref{sec:simulations} before we draw the conclusions in Section~\ref{sec:conc}. 

\emph{Notations:} Column vectors and matrices are given in lowercase and uppercase boldface letters, respectively. $\norm{\cdot}$, $\abs{\cdot}$, and $(\cdot)^\H$ denote respectively the Euclidean norm, absolute value, and Hermitian transpose. $\mat{I}$ is an identity matrix. Define the collection $\collec{a} := (a_1,\ldots,a_\abs{\setK})$. $\mathcal{CN}(\mat{0},\mat{A})$ denotes a circularly-symmetric Gaussian complex random vector with covariance matrix $\mat{A}$. 

%% file: sections/system_model.tex
\section{System Model and Problem Formulation}\label{sec:sys_model}
Consider a set of cells $\setK = \br{1,\ldots,K}$. The base station in a cell $k$ serves a set of users $\setU_k$. We assume that the base stations and the users use multiple antennas. Let $n_k$ and $m_{ki}$ be the number of antennas at base station $k$ and user $i$ in cell $k$, respectively. The flat fading channel matrix from a base station $j$ to a user $i \in \setU_k$ in cell $k$ is $\bH_{jki} \in \mathbb{C}^{m_{ki} \times n_j}$. 

The received signal at a user $i \in \setU_k$ is 
\begin{equation}
\mat{y}_{ki} = \sum \nolimits_{l \in \setK} \sum\nolimits_{j \in \setU_l}  \bH_{l k i} \mat{x}_{ji} + \bz_{ki},
\end{equation}
where the precoded signal $\mat{x}_{ji} \in \mathbb{C}^{n_j}$ has zero mean and covariance $\mathbb{E}\{\mat{x}_{ji}\mat{x}_{ji}^\H\} = \bX_{ji}$, and $\bz_{ki} \sim \mathcal{CN}(0,\bI \sigma^2)$ is additive white Gaussian noise. 

We assume that the maximum transmission power at a base station $k$ is restricted to $P_k$ and define the set of feasible transmission strategies of a base station $k$ as\footnote{More general linear power constraints can be also adopted similar to the model in \cite[Section II.A]{Bjornson2012}.}
\begin{multline}\label{eq:tx_set}
\setX_k = \left\{ (\bX_{k1}, \ldots, \bX_{k\abs{\setU_k}}) \mid \bX_{ki} \in \mathbb{C}^{n_k \times n_k}, \bX_{ki} \succeq \mat{0},  \right. \\ \left. \text{for all } i \in \setU_k, \sum\nolimits_{i \in \setU_k} \tr{\bX_{ki}} \leq P_k \right\} .
\end{multline}

Define the set of all feasible transmit covariance matrices of all base stations as 
\begin{equation}
\setX = \setX_1\times \cdots \times \setX_K.
\end{equation}

For a given $\bX \in \setX$ and assuming linear MMSE decoding, the achievable rate at user $i \in \setU_k$ is written as
\begin{equation}\label{eq:rate}
R_{ki}(\bX) = \log_2 \abs{\bI +  \bZ_{ki}^{-1}\bH_{kki} \bX_{ki} \bH_{kki}^\H},
\end{equation}
\noindent where the interference and noise covariance matrix is 
\begin{align}\nonumber
\bZ_{ki} = & \sigma^2 \bI +  \underbrace{\sum\limits_{l \in \setU_k\setminus\{i\}} \bH_{kki} \bX_{kl} \bH_{kki}^\H}_\text{intracell interference} \\ \label{eq:int_cov} 
& + \underbrace{\sum\limits_{j \in \setK\setminus \{k\}} \sum\limits_{l' \in \setU_j} \bH_{jki} \bX_{jl'} \bH_{jki}^\H}_\text{intercell interference}.
\end{align}

We are interested in finding the maximum sum rate operating point which is a solution of the following problem:
\begin{equation}\label{eq:original_prob0}
\maximize_{\bX} ~~ \sum_{k \in \setK} \sum_{i \in \setU_k}  {R_{ki}(\bX)} \quad s.t. ~~ \bX \in \setX.
\end{equation}
\noindent Problem \eqref{eq:original_prob0} is NP-hard \cite{Liu2011}.

%% file: sections/problem_reformulation.tex
\section{Problem Reformulation}\label{sec:Bender}
In this section, we will provide an equivalent formulation of the maximum sum rate problem in \eqref{eq:original_prob0}, which will enable the application of the generalized Benders decomposition. The reformulation is based on the following lemma.
\begin{lemma}\label{thm:eqv_rate}
Define
\begin{equation}\label{eq:rate2}
\underline{R}_{ki}(\bX, {\bY_{ki}}) = \log_2 \abs{{\bZ_{ki} +  \bH_{kki} \bX_{ki} \bH_{kki}^\H}} +  \log_2 \abs{\bY_{ki}}.
\end{equation}
Then, 
\begin{itemize}
\item[(i)] $R_{ki}(\bX) \geq \underline{R}_{ki}(\bX, {\bY_{ki}}) \text{ for } \mat{0} \prec\bY_{ki}$ and $\bY_{ki} \preceq \bZ_{ki}^{-1}$, with equality if $\bY_{ki}$ and $\bZ_{ki}^{-1}$ have the same eigenvalues. 
\item[(ii)] $\underline{R}_{ki}(\bX, {\bY_{ki}})$ is jointly concave in $\bX$ and $\bY_{ki}$.
\end{itemize}

\end{lemma}
\begin{IEEEproof}
The proof is provided in Appendix \ref{proof:eqv_rate}.
\end{IEEEproof}

We can formulate the following sum rate problem using the lower bound on the achievable rate in \eqref{eq:rate2} as
\begin{subequations}\label{eq:the_prob}
\begin{align}
\maximize_{\bX} \maximize_{\collec{\bY}} & ~~ \sum_{k \in \setK} \sum_{i \in \setU_k} \underline{R}_{ki}(\bX, \bY_{ki}) \\ \label{con:X}
s.t. & ~~ \bX \in \setX,\\ \label{con:BMI}
& ~~ \bZ_{ki} \bY_{ki} \preceq  \bI,  ~~ i \in \setU_k, k \in \setK,\\ \label{con:Y}
& ~~\bY_{ki} \succ \mat{0}, ~~ i \in \setU_k, k \in \setK.
\end{align}
\end{subequations}
By using Lemma \ref{thm:eqv_rate}, the equivalence of the Problem~\ref{eq:the_prob} and the original problem in \eqref{eq:original_prob0} can be established.

\begin{theorem}
Problem \eqref{eq:the_prob} and Problem \eqref{eq:original_prob0} are equivalent.
\end{theorem}
\begin{IEEEproof}
For any $\bX \in \setX$, an optimal value of $\bY_{ki}$ is $\bZ_{ki}^{-1}$ for all $i \in \setU_k, k \in \setK$ following (i) in Lemma \ref{thm:eqv_rate}. Accordingly, with $\bY_{ki}=\bZ_{ki}^{-1}$ for all $i \in \setU_k, k \in \setK$, the objective functions as well as the constraint sets of Problem \eqref{eq:the_prob} and Problem \eqref{eq:original_prob0} are identical.
\end{IEEEproof}

In Problem \eqref{eq:the_prob}, the objective is jointly concave in the optimization variables according to (ii) in Lemma \ref{thm:eqv_rate}, and the constraints \eqref{con:X} and \eqref{con:Y} are convex. The non convexity of Problem \eqref{eq:the_prob} is due to the bilinear matrix inequality constraints in \eqref{con:BMI}. Problem \eqref{eq:the_prob}, however, has the following significant property: If the variables $\collec{\bY}$ are fixed, then Problem~\eqref{eq:the_prob} is convex in $\bX$. Also, if we fix $\bX$, then the problem is convex in $\collec{\bY}$. For non convex problems with such attributes, a method called Benders decomposition has been proposed in \cite{Benders1962} for linear programs and generalized in \cite{Geoffrion72} to a wider class of problems utilizing nonlinear duality theory.

\subsection{Generalized Benders Decomposition}
The main idea of the generalized Benders decomposition is the projection of the problem onto the space of a specific set of variables \cite{Geoffrion72} leading to inner and outer optimization problems. In our case, the projection of Problem \eqref{eq:the_prob} in the space of the $\bX$ variables reformulates Problem \eqref{eq:the_prob} to
\begin{equation}\label{eq:the_feas_prob_proj}
\maximize_{\bX} ~~ v(\bX) \quad s.t. ~~\bX \in \setX \cap \setV,
\end{equation}
where $v(\bX)$ is defined by the inner optimization, also called the \underline{primal problem}:
\begin{subequations}\label{eq:primal}
\begin{align}
v(\bX) = \maximize_{\collec{\bY}} & ~~ \sum_{k \in \setK} \sum_{i \in \setU_k} \underline{R}_{ki}(\bX, \bY_{ki})\\ \label{eq:inner2} 
s.t. & ~~ \bZ_{ki} \bY_{ki} \preceq  \bI,  ~~ i \in \setU_k, k \in \setK,\\
& ~~ \bY_{ki} \succ \mat{0}, ~~ i \in \setU_k, k \in \setK,
\end{align}
\end{subequations}
and 
\begin{equation}\label{eq:setV}
\setV = \left\{ \bX \mid \bZ_{ki} \bY_{ki} \preceq \bI \textrm{ for some }\bY_{ki} \succ \mat{0}, i \in \setU_k, k \in \setK\right\}.
\end{equation}
The set $\setX \cap \setV$ in \eqref{eq:the_feas_prob_proj} represents the projection of the feasible region of Problem \eqref{eq:the_prob} onto the $\bX$ variable space. This is generally important in order to ensure feasibility of the primal problem in~\eqref{eq:primal}. However, in our case, the set $\setV$ in~\eqref{eq:setV} includes the set $\setX$ since it is always possible to find a feasible set of variables $\collec{\bY}$ for any $\bX \in \setX$. Hence, the primal problem in \eqref{eq:primal} is always feasible and we can consequently remove the constraint associated with $\setV$.

As Problem \eqref{eq:the_feas_prob_proj} is difficult to solve, the inner maximization problem in \eqref{eq:primal} is considered in its dual form. 

The Lagrangian of the primal problem in \eqref{eq:primal} is
\begin{multline}\label{eq:Lagrange}
\mathcal{L}(\bX,\collec{\bY},\collec{\mat{\Gamma}}) =  \\ \sum\limits_{k \in \setK} \sum\limits_{i \in \setU_k} \underline{R}_{ki}(\bX, \bY_{ki}) + \tr{\mat{\Gamma}_{ki} \pp{ \bI - \bZ_{ki} \bY_{ki}}},
\end{multline}
\noindent where $\collec{\mat{\Gamma}}$ are Lagrangian multipliers associated with constraints \eqref{eq:inner2}. Since strong duality holds for the inner maximization problem, Problem \eqref{eq:the_feas_prob_proj} can be reformulated as
\begin{subequations}\label{eq:relaxed_dual0}
\begin{align}
\maximize_{\bX} & ~~ v({\bX})\\
s.t. 	& ~~ v(\bX) = \minimize_{\collec{\mat{\Gamma}}} \maximize_{\collec{\bY}} ~~ \mathcal{L}(\bX,\collec{\bY}, \collec{\mat{\Gamma}}),\\ 
	& ~~ \qquad \qquad \quad ~~ s.t. ~~ \mat{\Gamma}_{ki} \succeq \mat{0}, ~~ i \in \setU_k, k \in \setK, \\
	& ~~ \qquad \qquad \qquad \quad ~~ \bY_{ki} \succeq \mat{0}, ~~ i \in \setU_k, k \in \setK,\\
	& ~~\bX \in \setX.
	\end{align}
\end{subequations}
By using the definition of a minimum, Problem~\eqref{eq:relaxed_dual0} can be reformulated to what will be called the \underline{master problem}:
\begin{subequations}\label{eq:relaxed_dual}
\begin{align}
\maximize_{\bX} & ~~ v({\bX})\\ \label{eq:inner_relaxed_dual1}
s.t. 	& ~~ v(\bX) \leq \maximize_{\collec{\bY}} ~~ \mathcal{L}(\bX,\collec{\bY}, \collec{\mat{\Gamma}}),\\ \nonumber
	& ~~ \qquad \qquad \qquad \textrm{for all } \mat{\Gamma}_{ki} \succeq \mat{0}, i \in \setU_k, k \in \setK, \\ \label{eq:inner_relaxed_dual3}
	& ~~ \qquad \qquad s.t. ~~ \bY_{ki} \succeq \mat{0}, ~~ i \in \setU_k, k \in \setK,\\
	& ~~\bX \in \setX.
	\end{align}
\end{subequations}
The form of Problem~\eqref{eq:relaxed_dual} is convenient for two optimization approaches which we address next.%

%% file: sections/upper_bound.tex
\section{Optimization Approaches}\label{sec:approaches}
Optimization methods which utilize the generalized Benders decomposition alternate between solving the primal problem in~\eqref{eq:primal} and the master problem in~\eqref{eq:relaxed_dual}. The solution of the primal problem is used as a lower bound on the maximum sum rate, while an upper bound is characterized by the master problem. However, the master problem in~\eqref{eq:relaxed_dual} is hard to solve due to the existence of the inner optimization problem \eqref{eq:inner_relaxed_dual1}-\eqref{eq:inner_relaxed_dual3}. Two methods which deal with this problem are presented in the next subsections.
\subsection{Approach 1}\label{sec:GOS}
This approach is based on the work in \cite{Floudas1989, Geoffrion72} and is described using Algorithm~\ref{alg1}. In each iteration $t$ of the algorithm two problems are solved. First in Line 3, for fixed variables $\bX^{t-1}$ (obtained from iteration $t-1$) the convex primal problem in \eqref{eq:primal} is solved to obtain $\collec{\bY^t}$ and a set of duals $\collec{\mat{\Gamma}^t}$ associated with constraints \eqref{eq:inner2}. The second optimization in Line 4 uses the solution of the primal problem as fixed inputs for the master problem problem in \eqref{eq:relaxed_dual} to obtain $\bX^{t}$. Notice that for fixed $\collec{\bY^t}$ and $\collec{\mat{\Gamma}^t}$, the master problem problem is convex in $\bX$.

In each iteration $t$ of Algorithm~\ref{alg1} the following is satisfied:
\begin{equation}\label{eq:property3}
\mathcal{L}(\bX^t,\collec{\bY^t}, \collec{\mat{\Gamma}^t}) \geq \sum\limits_{k \in \setK} \sum\limits_{i \in \setU_k} \underline{R}_{ki}(\bX^t, \bY_{ki}^t),
\end{equation}
whose proof follows similar lines as the proof in \cite[Appendix B]{Floudas1989}. That is, the solution of the master problem is always larger than the solution of the primal problem in Algorithm~\ref{alg1}. The convergence of Algorithm~\ref{alg1} to a local optimum is guaranteed according to \cite[Theorem 2.5]{Geoffrion72}.%
%
\begin{figure*}[t]
\begin{align} \tag{LIN}\label{eq:linearization} 
\mathcal{L}(\bX,\collec{\bY}, \collec{\mat{\Gamma}^t})\evat_{\collec{\bY^t}}^\lin &= \underbrace{\mathcal{L}(\bX,\collec{\bY^t},\collec{\mat{\Gamma}^t})}_{\text{in } \eqref{eq:Lagrange}} + \sum_{k\in \setK} \sum_{i\in \setU_k} \text{tr}\Big(\underbrace{\nabla_{{\bY_{ki}}} \mathcal{L}(\bX,\collec{\bY},\collec{\mat{\Gamma}^t})\evat_{\bY_{ki}^t}}_{=({\bY}_{ki}^t)^{-1} - \mat{\Gamma}_{ki}^t \bZ_{ki}}\Big)(\bY_{ki} - \bY_{ki}^t)\\ \tag{LIN-LB} \label{eq:lb_linearization}
{\mathcal{L}}(\bX,\collec{\bY},\collec{\mat{\Gamma}^t})\evat_{\collec{\bY^t}}^\lblin  &=  \mathcal{L}(\bX,\collec{\bY^t},\collec{\mat{\Gamma}^t}) - \sum_{k\in \setK} \sum_{i\in \setU_k} \frac{1}{2}\snorm{(\bY_{ki}^t)^{-1} - \mat{\Gamma}_{ki}^t\bZ_{ki} - (\bY_{ki} - \bY_{ki}^t)}
\end{align}
\hrulefill
\end{figure*}
\subsection{Approach 2}\label{sec:GOP}
\input{sections/algorithm1}
\input{sections/algorithm2}
Similar to Approach 1, the second approach relies on a solution for the master problem to obtain an upper bound on the sum rate. Based on the method in \cite{Floudas90aglobal}, a linearization of the Lagrangian $\mathcal{L}(\bX,\collec{\bY}, \collec{\mat{\Gamma}^t})$ is applied around $\collec{\bY^t}$, where, similar to Approach 1, $\collec{\bY^t}$ and $\collec{\mat{\Gamma}^t}$ are from the optimization of the primal problem in \eqref{eq:primal} at the $t$th iteration of the algorithm. The linearization of the Lagrangian corresponds to the first-order Taylor series expansion about $\collec{\bY^t}$ as given in \eqref{eq:linearization} at the top of the next page. This function satisfies the following \cite[Property 4]{Floudas90aglobal}:
\begin{equation}
\maximize_{\collec{\bY}} \mathcal{L}(\bX,\collec{\bY},\collec{\mat{\Gamma}^t}) \leq \maximize_{\collec{\bY}}  \mathcal{L}(\bX,\collec{\bY}, \collec{\mat{\Gamma}^t})\evat_{\collec{\bY^t}}^\lin.
\end{equation}
\noindent The above inequality ensures achieving an upper bound on the sum rate by the master problem \eqref{eq:relaxed_dual} with the Lagrangian replaced by the linearization. While in \cite{Floudas90aglobal} a global optimization framework is constructed based on the linearization approach, it proves hard to apply the existing techniques in our case where the variables are Hermitian matrices. Whenever a method can be found to maximize \eqref{eq:linearization} in $\bX$ and $\collec{\bY}$, a global optimization scheme can be constructed.

In this work, we will use an approximation on the linearization of the Lagrangian in \eqref{eq:linearization} with the following property.
\begin{proposition}\label{thm:lowerbound_lin}
For the expressions given in \eqref{eq:linearization} and \eqref{eq:lb_linearization} at the top of the next page, the following holds 
\begin{equation}
\maximize_{\collec{\bY}} \mathcal{L}(\bX,\collec{\bY},\collec{\mat{\Gamma}^t})\evat_{\collec{\bY^t}}^\lin \geq \maximize_{\collec{\bY}}\mathcal{L}(\bX,\collec{\bY},\collec{\mat{\Gamma}^t})\evat_{\collec{\bY^t}}^\lblin.
\end{equation}
\end{proposition}
\begin{IEEEproof}
The proof is provided in Appendix \ref{proof:lowerbound_lin}.
\end{IEEEproof}%

The algorithm for Approach 2 is given in Algorithm~\ref{alg2} and alternates between solving the primal problem in \eqref{eq:primal} to obtain a lower bound on the sum rate and solving the master problem in \eqref{eq:relaxed_dual_alg} to obtain an approximate upper bound on the sum rate. Note that \eqref{eq:lb_linearization} is jointly concave in $\bX$ and $\collec{\bY}$ and hence Problem \eqref{eq:relaxed_dual_alg} is convex. As in Algorithm~\ref{alg1}, solving the primal problem in Line 3 for fixed $\bX^{t-1}$ gives $\collec{\bY^t}$ and the duals $\collec{\mat{\Gamma}^t}$ associated with constraints \eqref{eq:inner2}. These parameters are used for solving \eqref{eq:relaxed_dual_alg} which is the master problem in \eqref{eq:relaxed_dual} with the Lagrangian replaced by \eqref{eq:lb_linearization}.

Although simulation results show ensured convergence of Algorithm~\ref{alg2}, a global convergence proof is hard to conduct mainly due to the lower bound property in Proposition \ref{thm:lowerbound_lin}.

\subsection{Complexity}

Using the same notation as in \cite{Shi2011}, let $\kappa$ be the total number of users in the network and $R$ and $T$ be the number of antennas at each receiver and transmitter, respectively. The complexity in each iteration of Algorithm~\ref{alg1} or Algorithm~\ref{alg2} is dominated by calculating the interference and noise covariance matrices in \eqref{eq:int_cov}, and solving the primal and master problems. 

As in the alternating weighted MMSE approach \cite{Shi2011}, the complexity of calculating $\bZ_{ki}$ in \eqref{eq:int_cov} for all receivers is $\mathcal{O}(\kappa^2TR^2)$. The solution of the primal problem in \eqref{eq:primal} is the inverse of $\bZ_{ki}$ and thus requires for all users $\mathcal{O}(\kappa R^3)$ arithmetic operations which is similar to the complexity for calculating the decoders in \cite{Shi2011}. The complexity for solving the master problem, which is a convex semidefinite program, is hard to quantify due to the dependency on the structure of the problem and its representation in the solver. Generally however, it is known that interior point methods for semidefinite programming are very efficient \cite{Vandenberghe1996}. Having that the calculation of the precoders in \cite{Shi2011} is a quadratically constrained quadratic program which can be solved by the more general semidefinite programs, we infer that our algorithm has higher complexity per iteration compared to the algorithm in \cite{Shi2011}.
%
%
%
%
%
%

%% file: sections/algorithm1.tex
\begin{algorithm}[t]
\caption{\label{alg1} Approach 1 for sum rate optimization.}
\begin{algorithmic}[1]
\Statex \textbf{Initilize}: $t = 0$, $\bX^0 \in \setX$, accuracy measure $\epsilon$
\Repeat
\State $t = t+1$;
\State \parbox[t]{\dimexpr\linewidth-\algorithmicindent}{Solve primal problem \eqref{eq:primal} for fixed $\bX^{t-1}$ to obtain $\bY^t_{ki}$ and $\mat{\Gamma}_{ki}^t$ for all $i \in \setU_k, k \in \setK$.\strut}
\State \parbox[t]{\dimexpr\linewidth-\algorithmicindent}{Solve master problem \eqref{eq:relaxed_dual} for fixed $\bY^t_{ki}$ and $\mat{\Gamma}_{ki}^t$ to obtain $\bX^{t}$.\strut}
\Until $\sum\nolimits_{k \in \setK} \sum\nolimits_{i \in \setU_k} ({R}_{ki}(\bX^t) - {R}_{ki}(\bX^{t-1})) < \epsilon$
\end{algorithmic} 
\end{algorithm}%

%% file: sections/algorithm2.tex
\begin{algorithm}[t]
\caption{\label{alg2} Approach 2 for sum rate optimization.}
\begin{algorithmic}[1]
\Statex \textbf{Initilize}: $t = 0$, $\bX^0 \in \setX$, accuracy measure $\epsilon$
\Repeat
\State $t = t+1$;
\State \parbox[t]{\dimexpr\linewidth-\algorithmicindent}{Solve primal problem \eqref{eq:primal} for fixed $\bX^{t-1}$ to obtain $\bY^t_{ki}$ and $\mat{\Gamma}_{ki}^t$ for all $i \in \setU_k, k \in \setK$.\strut}
\State Solve master problem \eqref{eq:relaxed_dual} using \eqref{eq:lb_linearization}
\begin{subequations}\label{eq:relaxed_dual_alg}
\begin{align}
\bX^t = \argmax_{\collec{\bY}, \bX \in \setX} & ~~ {\mathcal{L}}(\bX,\collec{\bY},\collec{ \mat{\Gamma}^t})\evat_{\collec{\bY^t}}^\lblin \\
s.t. 	& ~~\bY_{ki} \succ \mat{0}, ~~i \in \setU_k, k \in \setK.
	\end{align}
\end{subequations}
\Until $\sum\nolimits_{k \in \setK} \sum\nolimits_{i \in \setU_k} ({R}_{ki}(\bX^t) - {R}_{ki}(\bX^{t-1})) < \epsilon$
\end{algorithmic} 
\end{algorithm}%

%% file: sections/simulations.tex
%
\begin{figure}[t]
  \centering
 \includegraphics[width=\linewidth,clip]{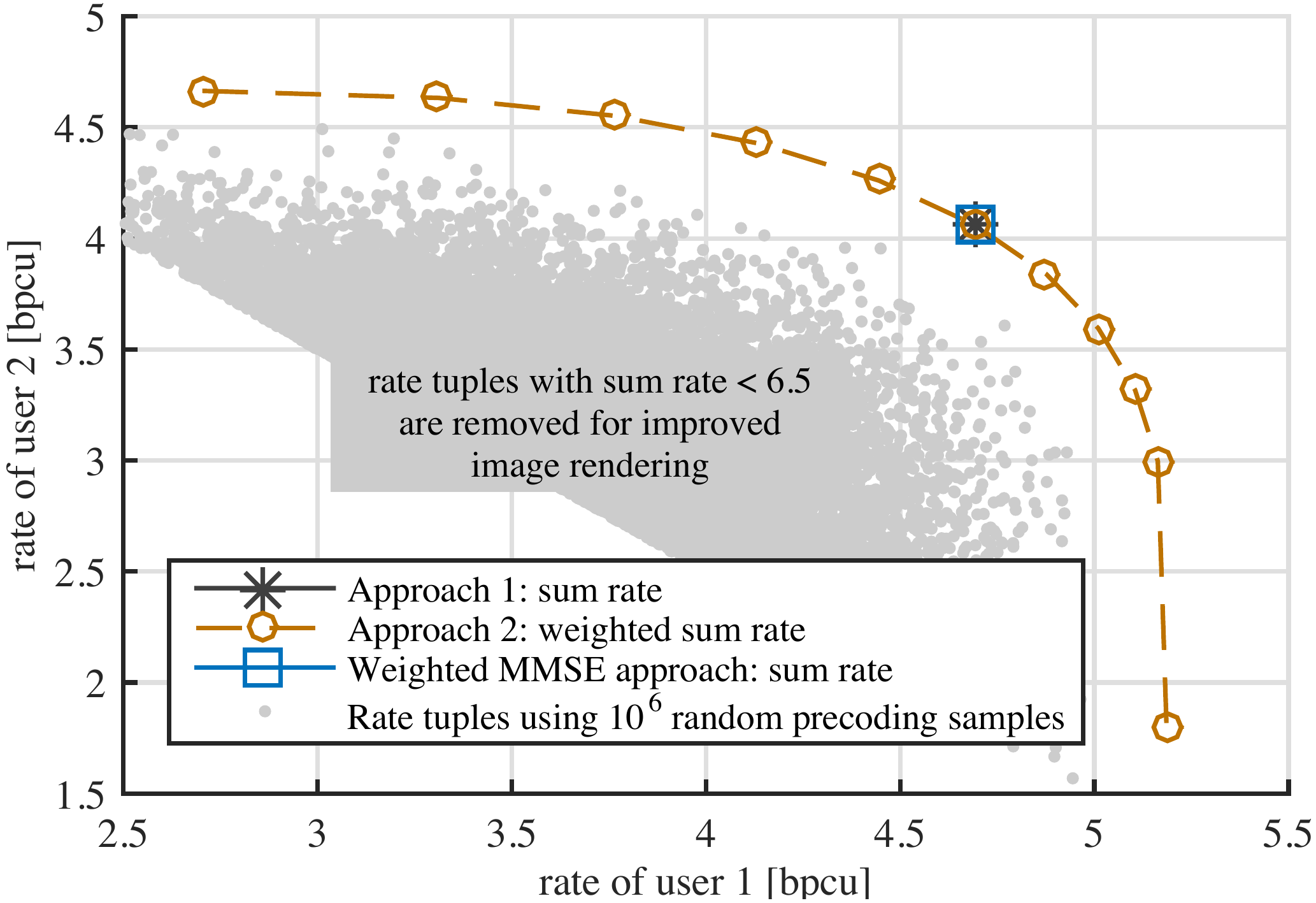}
  \caption{\label{fig:rate_region}Illustration of a two-user rate region. The number of antennas used at each transmitter and receiver is two.}
  \vspace{-0.25cm}
\end{figure}
\section{Illustrations}\label{sec:simulations}
In the simulation examples, we set the transmission power as $P_k = 1$ for all $k \in\setK$, and choose $\text{SNR} = 1/\sigma^2 = 10 \text{dB}$. The accuracy measure for the algorithms is set to $\epsilon = 10^{-5}$. We use CVX \cite{cvx} with the SeduMi solver \cite{SEDUMI} for solving the optimization problems within the algorithms.

In \figurename~\ref{fig:rate_region}, the achievable rate region in a setting with two cells and one user in each cell is illustrated. The cloud of points corresponds to $10^6$ randomly generated achievable rate tuples. Here, we use the algorithm proposed in \cite{Mittelbach2012} to generate the transmit covariance matrices with trace constraints satisfying \eqref{eq:tx_set}. The outer boundary of the rate region is plotted using Algorithm~\ref{alg2} where each point corresponds to a weighted sum rate optimization. The weighted performance optimization includes weighting factors for each term in the summation in the objective of \eqref{eq:original_prob0} which we have excluded in our model due to obvious extension possibility. In the two user case in \figurename~\ref{fig:rate_region}, ten weighting factors for user one $\lambda_1$ have been chosen uniformly from $[0,1]$ and the weights for users two are set as $\lambda_2 = 1 - \lambda_1$. It can be observed that the high number of randomly generated rate tuples is not sufficient to obtain points near to the plotted boundary. Hence, for benchmarking purposes, an exhaustive search on randomly generated rate tuples is not likely to be efficient especially when more than two antennas are used at the transmitters and the receivers as well as for larger number of users. In \figurename~\ref{fig:rate_region}, the optimized sum rate point is distinguished with a square and star marker and obtained by all three algorithms. Note however that non of the algorithms is guaranteed to converge to the global optimum of \eqref{eq:original_prob0} and currently it is not possible to verify whether the obtained point corresponds to the maximum sum rate.

\begin{figure}[t]
  \centering
 \includegraphics[width=\linewidth,clip]{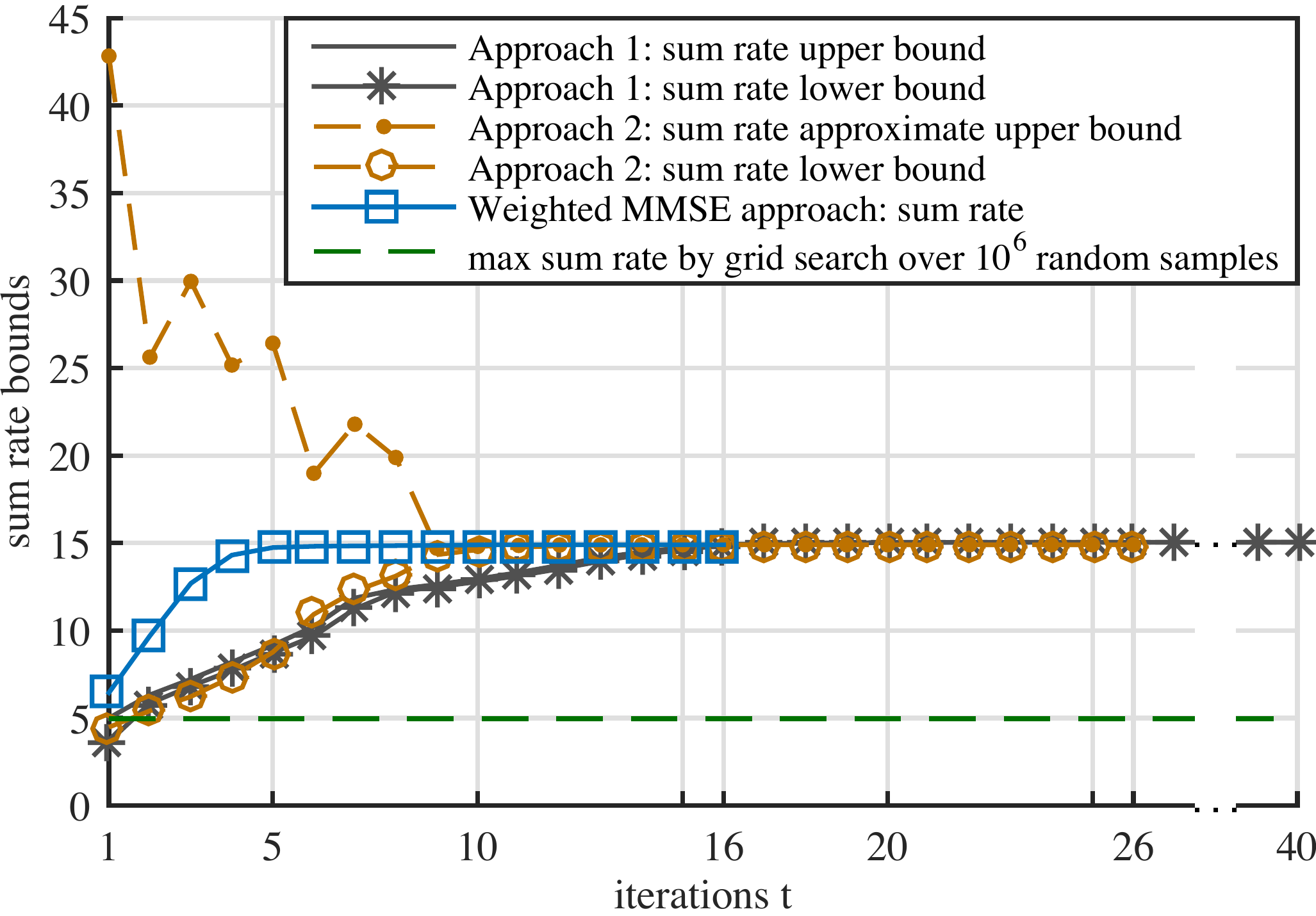}
  \caption{\label{fig:convergence} Convergence behavior for $K=10$ cells with one user per cell. The number of antennas at each transmitter and receiver is two.}
  \vspace{-0.25cm}
\end{figure}
In \figurename~\ref{fig:convergence}, the convergence behavior of our algorithms is shown for a setting with ten cells and one user in each cell. It can be observed that the upper bound approximation in Approach 2 (Algorithm~\ref{alg2}) obtained as the solution of \eqref{eq:relaxed_dual_alg} is not monotonically decreasing. The upper bound on the sum rate in Approach 1 (Algorithm~\ref{alg1}) is, in comparison to Approach 2, much closer to its associated lower bound. Despite this fact, Approach 1 needs higher number of iterations of $40$ according to the example in \figurename~\ref{fig:convergence}. The alternating weighted MMSE approach \cite{Shi2011} shows fast convergence behavior to the solution which is also obtained by the other algorithms. Grid search for the maximum sum rate over $10^6$ randomly generated achievable rate tuples is shown to be not adequate for benchmarking purposes.

%% file: sections/conclusions.tex
\section{Conclusions}\label{sec:conc}
Sum rate optimization in multicell MIMO is reformulated using generalized Benders decomposition. As a result, two optimization approaches are applied. Numerical results show comparable efficiency of the proposed methods to existing optimization methods. Future work will study the possibility of finding an exhaustive optimization method for the master problem with the linearization of the Lagrangian which could aid in constructing a global optimization method for finding the maximum sum rate operating point. 

%% file: sections/ack.tex
\section*{Acknowledgment}\label{sec:ack}
The authors would like to thank Rasmus Brandt for providing the implementation of the weighted MMSE approach. The authors would also like to thank Christoph Hellings for pointing out a mistake in an earlier version of the paper.

%% file: sections/appendix.tex
\appendix
\subsection{Proof of Lemma \ref{thm:eqv_rate}}\label{proof:eqv_rate}
In item (i), we have to show that $R_{ki}(\bX) - \underline{R}_{ki}(\bX, {\bY_{ki}}) = \log_2 \abs{\bZ_{ki}^{-1}} - \log_2 \abs{\bY_{ki}} \geq 0$. Note that the function $\log_2\abs{\mat{A}}$ with $\mat{A} \succ \mat{0}$ can be equivalently written as $\tr{\phi\pp{\mat{A}}}$ where $\phi$ is a matrix-monotone function \cite[Definition 3.1]{Jorswieck2007} using similar derivation steps as \cite[Example 3.2]{Jorswieck2007}. We then have $\log_2 \abs{\bY_{ki}} \leq \log_2 \abs{\bZ_{ki}^{-1}}$ for $\mat{0} \prec\bY_{ki} \preceq \bZ_{ki}^{-1}$ following the definition of matrix-monotone functions in \cite[Definition 3.2]{Jorswieck2007}. Since a matrix-monotone function $\phi\pp{\mat{A}}$ is affected only by the eigenvalues of the matrix $\mat{A}$ \cite[Definition 3.1]{Jorswieck2007}, the equality in item (i) is achieved when the eigenvalues of $\bY_{ki}$ and $\bZ_{ki}^{-1}$ are the same. The proof of item (ii) follows from \cite[Lemma 3.10]{Jorswieck2007} which states that the trace of a matrix monotone function is concave and monotone.

\subsection{Proof of Proposition \ref{thm:lowerbound_lin}}\label{proof:lowerbound_lin}
Define $\mat{A}:=(\bY_{ki}^t)^{-1} - \mat{\Gamma}_{ki}^t \bZ_{ki}$ and $\mat{B}:=\bY_{ki} - \bY_{ki}^t$. Then
\begin{align}
\tr{\mat{A}\mat{B}} &=  \tr{\mat{A}\mat{B}} - \frac{\tr{\mat{A}^2 + \mat{B}^2}}{2} + \frac{\tr{\mat{A}^2 + \mat{B}^2}}{2},\\
& = - \frac{1}{2}\tr{\pp{\mat{A} - \mat{B}}^2} + \frac{1}{2}\tr{\mat{A}^2 + \mat{B}^2},\\ \label{eq:last_inequality}
& \geq - \frac{1}{2}\tr{\pp{\mat{A} - \mat{B}}^2} = - \frac{1}{2}\snorm{\mat{A} - \mat{B}}.
\end{align}